\documentclass[smallextendid]{svjour3wide}
\usepackage{amsmath,amssymb,bm}
\usepackage[dvips]{graphicx}
\usepackage{yfonts}
\usepackage{color}

\newcommand{\pset}{\mathcal{X}}
\newcommand{\psetUV}{\mathcal{X}_B}

\newcommand{\af}{i\tilde{h}}
\newcommand{\eaf}{i\tilde{h}^<}
\newcommand{\eh}{h^<}
\newcommand{\exf}{J}
\newcommand{\aexf}{\tilde{J}}

\begin{document}


\title{The most effective model for describing the universal behavior of unstable surface growth}
\titlerunning{The most effective model for the universal behavior of unstable surface growth}

\author{Yuki Minami \and Shin-ichi Sasa}

\institute{Y.~Minami \at National Institute of Advanced Industrial  Science and Technology (AIST), Ibaraki 305-8560, Japan \\ \email{minami-fehsf5@aist.go.jp}
\and S. Sasa \at Department of Physics, Kyoto University, Kyoto 606-8502, Japan \\ \email{sasa@scphys.kyoto-u.ac.jp} }

\date{\today}

\maketitle

\begin{abstract}
  We study a noisy Kuramoto-Sivashinsky (KS) equation which describes
  unstable surface growth and chemical turbulence. It has been conjectured
  that the universal long-wavelength behavior of the equation, which is
  characterized by scale-dependent parameters, is
  described by a Kardar-Parisi-Zhang (KPZ) equation. We consider
  this conjecture by analyzing a renormalization-group equation for
  a class of generalized  KPZ equations. We then uniquely determine
  the parameter values of the KPZ equation that most effectively  describes
   the universal long-wavelength behavior of the noisy KS equation.
\end{abstract}



\section{Introduction}
Eddy viscosity in turbulence, which can explain
how a vortex pattern emerges in a non-uniform turbulent
flow, depends on the observed length scales \cite{Frisch}.
As exemplified by the Richardson law \cite{Richardson},
there are cases in which a parameter of a macroscopic description is not
given as a definite value, but is rather expressed as a function
of the length scale. Another example of scale-dependent parameters
has been observed in one- or two- dimensional fluid dynamics, where the viscosity is
not uniquely defined in the hydrodynamic description \cite{review}.
Here, it seems reasonable to expect that such scale-dependent
parameters in a macroscopic description can be reproduced by
an effective stochastic system \cite{PhysRevA.16.732,Dominicis,Fournier-Frisch1978,Fournier-Frisch1983,Yakhot-Orszag,Yakhot-Orszag2,Yakhot-Smith,Eyink}.
In this paper,  we attempt to
determine the effective stochastic system  theoretically when scale-dependent
parameters are observed.


As the simplest example for scale-dependent parameters,
we consider the one-dimensional Kardar-Parisi-Zhang (KPZ) equation
\cite{PhysRevLett.56.889}.
It is known that the effective surface tension $\nu(\Lambda)$ at a
scale $2\pi/\Lambda$ for the equation is $\nu(\Lambda) = C_\nu \Lambda^{-1/2}$
in the limit $\Lambda \to 0$, which is similar to the Richardson law
for turbulence.
Recently, the KPZ equation was rigorously derived from a stochastic many-particle
model \cite{Bertini-Giacomin,PhysRevLett.104.230602}, and the so-called
KPZ class has been extensively discussed both theoretically and
experimentally \cite{Takeuchi-Sano2010,Takeuchi-Sano-Sasamoto-Spohn,Amir-Corwin-Quastel,Calabrese-Doussal,Imamura-Sasamoto,Hairer}.
However, in general, even if we find systems that may exhibit
scale-dependent parameters  similar to those for the  KPZ class,
a method to determine the parameter values of the corresponding KPZ equation has not yet been reported.


Specifically, let us consider a noisy Kuramoto-Sivashinsky (KS)
equation, which exhibits spatially extended chaos in the noiseless
limit \cite{kuramoto1976persistent,sivashinsky1977nonlinear,Kuramoto-book}.
The model describes turbulent chemical waves and unstable interface motions,
which are caused by negative surface tension.
It has been conjectured that a KPZ equation may be an effective model
 for describing the long-wavelength
behavior of the noisy KS equation; this conjecture is
referred to as the {\it Yakhot conjecture}
\cite{PhysRevA.24.642,PhysRevLett.60.1840}.
Indeed,  direct numerical simulations showed that
statistical properties of the  long wave length modes are similar
to those of the KPZ equations \cite{zaleski1989stochastic,PhysRevA.46.R7351,PhysRevE.47.911,sakaguchi2002effective,PhysRevE.71.046138}.
  
Here, one may recall the renormalization group (RG), which
is a standard method for studying  scale-dependent
parameters. For a given noisy KS equation, the RG equation
was calculated using a perturbation
theory \cite{PhysRevE.71.046138,Cuerno-Lauritsen}.
The infrared fixed point
of the RG equation determines the scale-dependent behavior
$\nu(\Lambda)=C_\nu \Lambda^{-1/2}$ in the limit $\Lambda \to 0$,
which has the same power-law form as that for the KPZ equations.
Nevertheless,
as shown below, the analysis at the infrared fixed point
of the RG equation cannot determine the parameter values of the corresponding
KPZ equation.


In this paper, we present a framework for studying the
effective description. We study an RG equation for generalized
KPZ equations that include noisy KS equations and KPZ equations.
We then consider solution trajectories of the RG equation, in which
each point flows to the infrared fixed point of the noisy KS equation
we study. The solution trajectories also approach a subspace
in the ultraviolet limit, which enables us to define a collection
of bare parameters of the generalized KPZ equations.
By using the lowest perturbation theory for the RG equation,
we uniquely determine the most effective model among such KPZ equations that
describes the infrared universal behavior of a noisy KS equation in the most efficient manner.

This paper is organized as follows.
In Sect.~\ref{sec:setup}, we introduce a class of models we study,  define scale-dependent parameters for the models, and review RG equations for the parameters. We also discuss ultraviolet and infrared behaviors of solution trajectories for the RG equation, and classify universal and non-universal properties of those. 
After that, we give a definition of  ``the most effective model''.
In Sect.~\ref{sec:parameterspace}, we simplify a representation of the trajectories so as to determine the most effective model. 
The solution trajectories for the RG equation are expressed as curves
in a five-dimensional parameter space.
Then, the trajectory for the noisy KS equation is attracted to a two-dimensional subspace, due to emergence of a time-reversal symmetry. We define the most effective model for the noisy KS equation in this subspace.
In Sect.~\ref{sec:MEmodel}, we determine parameter values of the most effective model from its definition.
In Sect.~\ref{sec:concluding}, we provide concluding remarks.
We discuss renormalizability of the KPZ equation and its relevant parameters.
We also remark an another application of our formalism to turbulence.  
In Appendix~\ref{sec:WTid}, we derive Ward-Takahashi identities for scale-dependent parameters from symmetries of our model.


\section{Setup}
\label{sec:setup}
We study models for the stochastic growth of a surface.
We assume that the time evolution of the height $h(x,t)$
of the surface is described by a generalized KPZ equation:
\begin{align}
\partial_t h &= \nu \partial_x^2 h
- K \partial_x^4 h
+\frac{\lambda}{2}(\partial_x h)^2 + \eta, \\
\langle \eta(x,t) \eta(x^{'},t^{'}) \rangle &=
  2(D-D_d \partial_x^2)\delta(t-t^{'})\delta(x-x^{'}),
\label{eq:gKPZ}
\end{align}
where $\nu$ is the surface tension, $K$ is the surface diffusion constant,
$\lambda$ is the strength of the non-linearity, and
$\eta(x,t)$ is the white noise.
Here, $D$ and $D_d$ are the strength of the noise.
When $K=D_d=0$, (\ref{eq:gKPZ}) is the  KPZ equation,
while when $D=D_d=0$, (\ref{eq:gKPZ}) with $\nu <0$
is the deterministic KS equation. We refer to (\ref{eq:gKPZ}) with $\nu < 0$
and $K, D, D_d > 0$, as the noisy KS equation.


The five parameters in (\ref{eq:gKPZ}) are collectively denoted
by $\pset \equiv (\nu, K, D, D_d, \lambda)$. More precisely,
these parameters are defined for a field $h(x)$ whose
Fourier transform $\hat{h}(k)$ is assumed to be zero for $|k| > \Lambda$.
$\Lambda$ is called a cut-off wavenumber.
We explicitly express the cutoff
dependence of the parameters as  $\pset (\Lambda)$. Here,
for a given model with $\pset (\Lambda_0)$, we define a model
with $\pset (\Lambda)$ for $\Lambda < \Lambda_0 $ by eliminating
the contribution $\Lambda \leq |k| \leq \Lambda_0$ in the dynamics,
which may be formally expressed as $\pset (\Lambda; \pset (\Lambda_0))$.
Note that we do not employ a rescaling transformation after the coarse-graining.
This functional relation trivially satisfies
\begin{equation}
\pset (\Lambda'; \pset (\Lambda_0))=
\pset (\Lambda'; \pset (\Lambda; \pset (\Lambda_0))).
\end{equation}
From this, we obtain the RG equation
\begin{align}
-\Lambda \frac{d \pset}{d \Lambda}=\Psi_\pset(\pset),
\label{eq:RGeq}
\end{align}
which determines $\pset (\Lambda; \pset (\Lambda_0))$
under an initial condition  $\pset (\Lambda_0)=\pset_0 $.
In the next subsection, we review  the RG equation
for the  generalized KPZ equations.

\subsection{definition of scale-dependent parameters}
We first define the scale-dependent parameters $\nu(\Lambda)$,
$D(\Lambda)$, $K(\Lambda)$, $D_d(\Lambda)$ and $\lambda(\Lambda)$,
and then introduce a perturbation theory leading to the equation for determining them.

We start with the generating functional $Z[\exf,\aexf]$ by which all statistical quantities of the KPZ equations are determined.
Following the Martin-Siggia-Rose-Janssen-deDominicis (MSRJD) formalism \cite{MSR,J,D1,D2},  $Z[\exf,\aexf]$ is expressed as
\begin{align}
  Z[\exf, \aexf] &=  \int \mathcal{D}[ h, \af] \exp\biggl[-S[h,\af;\Lambda_0] \nonumber \\
  &+\int_{-\infty}^{\infty} d\omega \int_{-\Lambda_0}^{\Lambda_0} dk \biggl(\exf(k,\omega) h(-k,-\omega)+ \aexf(k,\omega)  \af(-k,-\omega) \biggr) \biggr],
  \label{eq:generatingfunctional}
\end{align}
where $\af$ is the auxiliary field, $\exf$ and $\aexf$ are source fields,
and $S[h,\af;\Lambda_0]$ is the MSRJD action for the generalized KPZ equation.
Hereafter, we use the notation $A(k,\omega)$ for the Fourier transform of $A(x,t)$ for any field A.
The action $S[h,\af;\Lambda_0]$ is explicitly written as
\begin{align}
  S[h,\af;\Lambda_0]
  =& \frac{1}{2}\int_{-\infty}^{\infty}
  \frac{d\omega}{2\pi} \int^{\Lambda_0}_{-\Lambda_0}\frac{d k}{2\pi}
\begin{pmatrix}
h(-k,-\omega)  & \af(-k,-\omega)
\end{pmatrix}
G_0^{-1}(k,\omega)
\begin{pmatrix}
h(k,\omega)  \\ \af(k,\omega)
\end{pmatrix}
  \nonumber \\
   & +\frac{\lambda_0}{2} \int_{-\infty}^{\infty}
   \frac{d\omega_1 d\omega_2}{(2\pi)^2}
   \int^{\Lambda_0}_{-\Lambda_0}\frac{d k_1 d k_2}{(2\pi)^2}
   k_1 k_2\af( -k_1-k_2,-\omega_1-\omega_2) h(k_1, \omega_1)h(k_2,\omega_2),
\label{eq:action}
\end{align}
where  $G_0^{-1}$ is the inverse matrix of the bare propagator
\begin{align}
G_0^{-1}(k,\omega)=
\begin{pmatrix}
0 &  i\omega+\nu_0 k^2 +K_0 k^4 \\
-i\omega+\nu_0 k^2 +K_0 k^4  &-2(D_0 +D_{d0} k^2)
\end{pmatrix}.
\end{align}


Here, we consider a coarse-grained description
at a cutoff $\Lambda < \Lambda_0$.  Let us define
\begin{eqnarray}
  A^<(k, \omega) &\equiv&  \theta(\Lambda-k) A(k, \omega), \\
  A^>(k, \omega) &\equiv&  \theta(k-\Lambda) A(k, \omega),
\end{eqnarray}
for any quantity $A(k,\omega)$, where $\theta(x)$ is
the Heaviside step function. The statistical quantities
of $h^<$ are described by  the generating functional
$Z[\exf^<, \aexf^<] $ with replacement of $(\exf, \aexf)$ by
$(\exf^<, \aexf^<)$.  We thus define the  effective
MSRJD action $S[h^<,\af^<;\Lambda]$ by the relation
\begin{align}
  Z[\exf^<,\aexf^<]= &\int \mathcal{D}[ h^<, \af^<] 
  \exp\biggl[-S[h^<,\af^<;\Lambda]\nonumber \\
  &+\int_{-\infty}^{\infty} d\omega \int_{-\Lambda}^{\Lambda} dk \biggl(\exf^<(k,\omega) h^<(-k,-\omega)+ \aexf^<(k,\omega)  \af^<(-k,-\omega) \biggr) \biggr].
\end{align}
We can then confirm that $S[h^<,\af^<;\Lambda]$ is determined as
\begin{align}
 \exp \biggl[ -S[h^<,\af^<; \Lambda] \biggr]
 = \int \mathcal{D} [h^>, \af^>]
\exp \biggl[-S[h^<+h^>,\af^<+\af^>;\Lambda_0]\biggr].
\end{align}
Then, the propagator and the three point vertex function for
the effective MSRJD action at $\Lambda$ are defined as
\begin{align}
  (G^{-1})_{\tilde{h}h}(k_1, \omega_1; \Lambda)
  \delta(\omega_1+\omega_2)&\delta(k_1+k_2)
  \equiv \left. \frac{\delta^2 S[\eh,\eaf; \Lambda]}{\delta(\af(k_1, \omega_1))\delta(\eh(k_2, \omega_2))} \right|_{\eh,\eaf=0},
  \label{eq:def of propagator}\\
(G^{-1})_{\tilde{h}\tilde{h}}(k_1, \omega_1; \Lambda) \delta(\omega_1+\omega_2)&\delta(k_1+k_2)
  \equiv \left. \frac{\delta^2 S[\eh,\eaf; \Lambda]}{\delta(\eaf(k_1, \omega_1))\delta(\eaf(k_2, \omega_2))}\right|_{\eh,\eaf=0},
  \\
\Gamma_{\tilde{h} h h}(k_1, \omega_1;k_2,\omega_2; \Lambda)&\delta(\omega_1+\omega_2+\omega_3)\delta(k_1+k_2+k_3) 
\nonumber \\
 &\equiv \left.\frac{\delta^3 S[\eh,\eaf; \Lambda]}{\delta(\eaf(k_1, \omega_1))\delta(\eh(k_2, \omega_2))\delta(\eh(k_3,\omega_3))}\right|_{\eh,\eaf=0}.
 \label{eq:def of vertex}
\end{align}
From these quantities, we define the parameters as
\begin{align}
\nu(\Lambda)
  &\equiv \lim_{\omega, k \to 0}
   \frac{1}{2!}\frac{\partial^2 (G^{-1})_{\tilde{h}h}(k, \omega; \Lambda)}{\partial k^2},
   \label{eq:definition of nu} \\
 K(\Lambda)  &\equiv \lim_{\omega, k \to 0}\frac{1}{4!}\frac{\partial^4   (G^{-1})_{\tilde{h}h}(k, \omega; \Lambda)}{\partial k^4},
  \label{eq:definition of K} \\
-2D(\Lambda)  &\equiv \lim_{\omega, k \to 0} (G^{-1})_{\tilde{h}\tilde{h}}(k, \omega;\Lambda),
\label{eq:definition of D}\\
-2D_d(\Lambda)  &\equiv \lim_{\omega, k \to 0}\frac{1}{2!}\frac{\partial^2 (G^{-1})_{\tilde{h}\tilde{h}}(k, \omega; \Lambda)}{\partial k^2},
\label{eq:definition of Dd} \\
\lambda(\Lambda) & \equiv \lim_{\omega_1,\omega_2, k_1,k_2 \to 0} \frac{\partial^2 \Gamma_{\tilde{h} h h}(k_1, \omega_1;k_2,\omega_2;\Lambda)}{\partial k_1 \partial k_2}.
\label{eq:definition of lambda}
\end{align}
From a tilt symmetry of the generalized KPZ equation, we can obtain
\begin{align}
\lambda(\Lambda) = \lambda_0. \label{eq:lambda0}
\end{align}
In Appendix \ref{sec:WTid}, we will provide a non-perturbative proof for (\ref{eq:lambda0}) based on symmetry properties \cite{freyPRE1994,Canet2011}.
Below, we derive a set of equations that determines
$\nu(\Lambda)$, $D(\Lambda)$, $K(\Lambda)$, and $D_d(\Lambda)$.

\subsection{renormlization group equations}
We can calculate $(G^{-1})_{ij}(k,\omega; \Lambda)$ by using the perturbation
theory in $\lambda_0$. At the second-order level, the propagators
are calculated as
\begin{align}
(G^{-1})_{\tilde{h}h}(k, \omega; \Lambda) =&(G_0^{-1})_{\tilde{h}h} (k,\omega) \nonumber \\
 &+\lambda_0^2\int^{\infty}_{-\infty} \frac{d\Omega}{2\pi} \int_{ \Lambda \leq \vert q\vert \leq \Lambda_0 } \frac{d q}{2\pi}
\biggl[ kq(k-q)^2 (G_0)_{\tilde{h}h}(q,\Omega) C_0(k-q,\omega-\Omega)
\nonumber \\
&+ k q^2(k-q) (G_0)_{\tilde{h}h}(k-q,\omega-\Omega) C_0(q,\Omega) \biggr], \label{eq:propagator1}\\
(G^{-1})_{\tilde{h}\tilde{h}}(k, \omega; \Lambda) =&(G_0^{-1})_{\tilde{h}\tilde{h}}(k,\omega) \nonumber \\
&-2\lambda_0^2\int^{\infty}_{-\infty} \frac{d\Omega}{2\pi} \int_{ \Lambda \leq \vert q\vert \leq \Lambda_0 } \frac{d q}{2\pi} q^2(k-q)^2
 C_0(q,\Omega) C_0(k-q,\omega-\Omega),   \label{eq:propagator2}
\end{align}
where $C_0(k,\omega )$ is the bare correlation function defined by
\begin{align}
C_0(k,\omega) \equiv  2 ( D_0 +D_{d0} k^2) \vert (G_0)_{\tilde{h}h} (k, \omega) \vert^2.
\end{align}
In the calculation of (\ref{eq:propagator1}), one should carefully note the relation \cite{PhysRevE.71.046138}
\begin{align}
 \int_{ \Lambda \leq \vert q\vert \leq \Lambda_0 } \frac{d q}{2\pi} &q(k-q)^2 (G_0)_{\tilde{h}h}(q,\Omega) C_0(k-q,\omega-\Omega) \nonumber \\ 
&\neq  
\int_{ \Lambda \leq \vert q\vert \leq \Lambda_0 } \frac{d q}{2\pi}  q^2 (k-q) (G_0)_{\tilde{h}h}(k-q,\omega-\Omega) C_0(q,\Omega).
\end{align}
We emphasize that the Feynman rule does not distinguish these.

By setting $(\Lambda-\Lambda_0) /\Lambda_0 \ll 1$ for (\ref{eq:definition of nu}) - (\ref{eq:definition of Dd}), (\ref{eq:propagator1}) and (\ref{eq:propagator2}), we obtain the RG equation as
 \begin{align}
 -\Lambda \frac{d \nu(\Lambda)}{d \Lambda} &=\nu(\Lambda)\biggl[
    \frac{G}{F(1+F)^3}\biggl(3+F+(1-F)\frac{H}{G}\biggr)\biggr],
    \label{eq:RGnu}\\
 -\Lambda\frac{d K(\Lambda)}{d \Lambda} &=K(\Lambda)\biggl[
    \frac{G}{2(1+F)^5}\biggl(26-F+2F^2+F^3+(2-21F+6F^2+F^3)\frac{H}{G}\biggr)\biggr],
    \label{eq:RGK}\\
 -\Lambda\frac{d D(\Lambda)}{d \Lambda} &=D(\Lambda)\biggl[
    \frac{G}{(1+F)^3}\biggl(1+\frac{H}{G}\biggr)^2\biggr],
    \label{eq:RGD}\\
 -\Lambda \frac{d D_d(\Lambda)}{d \Lambda} &=D_d(\Lambda)\biggl[
    \frac{G^2}{2H(1+F)^5}\biggl(16+3F+F^2+2(9-5F)\frac{H}{G}
    +(2-13F-F^2)\frac{H^2}{G^2}\biggr)\biggr]
    \label{eq:RGDd}, 
 \end{align}
 where we have introduced the dimensionless parameters $F$, $G$ and $H$ as
 \begin{align}
 F&=\frac{\nu(\Lambda) }{K(\Lambda) \Lambda^2}, \\
 G&=\frac{\lambda_0^2 D(\Lambda)}{4\pi K^3(\Lambda) \Lambda^7}, \\
 H&=\frac{\lambda_0^2 D_d(\Lambda)}{4\pi K^3(\Lambda) \Lambda^5}. \label{eq:FGH}
 \end{align}

Here,  from (\ref{eq:RGnu})-(\ref{eq:RGDd}), we derive the autonomous
equation for $(F(\Lambda), G(\Lambda), H(\Lambda))$ as
\begin{align}
-\Lambda \frac{d F}{d \Lambda} =2F+\frac{G}{2(1+F)^5}\biggl[&
  6-12F+11F^2-F^4+(2+19F^2-8F^3-F^4)\frac{H}{G}\biggr],
  \label{eq:RGF}\\
-\Lambda \frac{d G}{d \Lambda} =7G-\frac{G^2}{2(1+F)^5}\biggl[&
  76-7F+4F^2+3F^3+(2-71F+14F^2+3F^3)\frac{H}{G} \nonumber \\
  &-2(1+F)^2\frac{H^2}{G^2}\biggr],
  \label{eq:RGG}\\
-\Lambda \frac{d H}{d \Lambda} =5H+\frac{G^2}{2(1+F)^5}\biggl[&
  16+3F+F^2-(60+7F+6F^2+3F^3)\frac{H}{G} \nonumber \\
  &-(4-50F+19F^2+3F^3)\frac{H^2}{G^2}\biggr].
  \label{eq:RGH}
\end{align}

\subsection{Infrared and ultraviolet behaviors of solution trajectories of the RG equation}
The stable fixed point of the equations (\ref{eq:RGF}) - (\ref{eq:RGH}) is found to be
$(F^*,G,^*,H^*)=(10.7593, \; 680.652,\;63.2614)$.
By substituting the fixed point values to (\ref{eq:RGnu})-(\ref{eq:RGDd}) and solving them, we obtain the scaling laws
\begin{align}
\nu(\Lambda)&=C_\nu\Lambda^{-0.5}, \label{eq:scaling nu}\\
D(\Lambda)&=C_D\Lambda^{-0.5}, \\
K(\Lambda)&=C_K\Lambda^{-2.5}, \\
D_d(\Lambda)&=C_{D_d}\Lambda^{-2.5}
\label{eq:scaling Dd},
\end{align}
where $C_\nu$, $C_D$, $C_K$, and $C_{D_d}$ are
constants that depend on the initial
condition $\pset_0$.

We next consider the dimensionless quantities given by
\begin{align}
\frac{1}{F}&=\frac{K(\Lambda) \Lambda^2}{\nu(\Lambda)},   \\
 \frac{H}{G}&=\frac{D_d(\Lambda)\Lambda^2}{D(\Lambda)}.
\end{align}
Substituting the scaling relations (\ref{eq:scaling nu}) - (\ref{eq:scaling Dd}) to these equalities, we have
\begin{align}
\frac{1}{F^*}&=\frac{C_K}{C_\nu}, \nonumber \\
\frac{H^*}{G^*}&=\frac{C_{D_d}}{C_D}.
\end{align}
Since $(F,G,H)$ takes the value $(10.7593, \; 680.652,\;63.2614)$
in the limit $\Lambda \rightarrow 0$,
we obtain
\begin{align}
\frac{C_K}{C_\nu}=\frac{C_{D_d}}{C_D}=0.0929,
\end{align}
which is independent of $\pset_0$
.
The singular behavior $\nu(\Lambda)=C_\nu\Lambda^{-1/2}$
implies that the effective surface tension depends on the observed scale $\Lambda$.
This is contrasted
with cases in which each $\pset(\Lambda)$ converges to a finite value in the limit $\Lambda \rightarrow 0$.
Then, $\pset(\Lambda=0)$ is interpreted as renormalized parameters measured
in experiments. Since the exponents characterizing
the divergent behaviors are common to all the models given by (\ref{eq:gKPZ}),
we refer to the power-law region as the universal range.
The smallest characteristic wavenumber scale is also denoted
by $\Lambda_{IR}$,  the value of which depends on $\pset_0$.
Then, the universal range is defined as $\Lambda \ll
\Lambda_{IR}$. As another common aspect of the RG equation
(\ref{eq:RGeq}), we observe  that  $\pset(\Lambda)$ shows a plateau region
in the ultraviolet limit when $\Lambda_0$ is sufficiently large.
This enables us to define a collection of bare parameters, which is
denoted by $\psetUV$.

\begin{figure}[htbp]
  \begin{center}
    \begin{tabular}{c}

      \begin{minipage}{0.5\hsize}
        \begin{center}
          \includegraphics[width=\hsize]{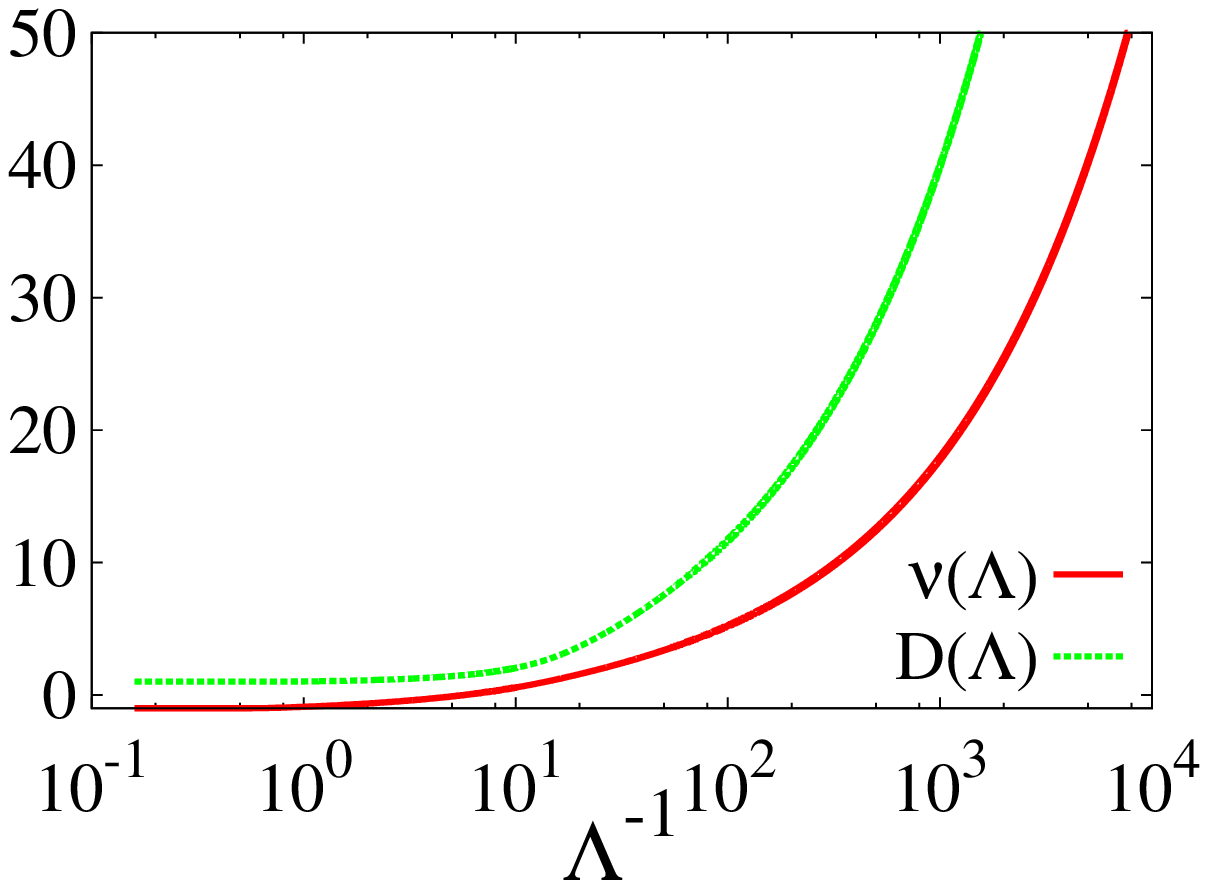}
        \end{center}
      \end{minipage}
 
      \begin{minipage}{0.5\hsize}
        \begin{center}
          \includegraphics[width=\hsize]{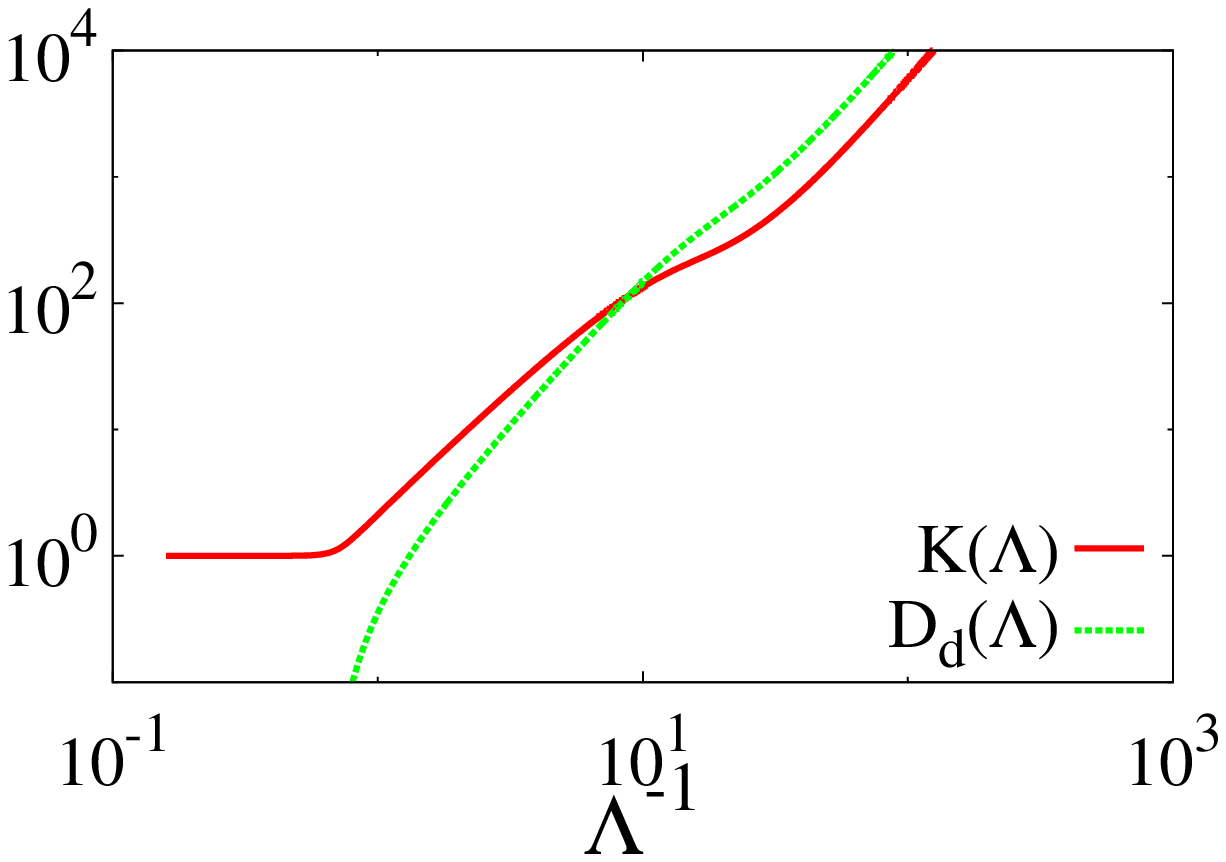}
        \end{center}
      \end{minipage}

    \end{tabular}
    \caption{Graphs of  $\nu(\Lambda)$, $D(\Lambda)$, $K(\Lambda)$, and $D_d(\Lambda)$ for $\pset_0^{KS}$.}
    \label{fig:nuD-KS}
  \end{center}
\end{figure}


Here, we focus on a specific model, a noisy KS equation with
$\pset_0^{KS}, (\nu_0=-1.0, D_0=0, K_0=D_{d0}=\lambda_0=1.0)$,  defined
at $\Lambda_0=2 \pi$. In Fig. \ref{fig:nuD-KS}, we display
the numerical solution of (\ref{eq:RGeq}) for this initial
condition $\pset_0^{KS}$. It can be seen that $\Lambda_0$ is in the
plateau region. Thus, the collection of the bare parameters
$\psetUV^{KS}$ is assumed to be identical to the initial condition
$\pset_0^{KS}$ without loss of accuracy.
On the other hand, the numerical solution in the infrared limit
obeys $\nu(\Lambda)=C_\nu\Lambda^{-0.5}$ and $D(\Lambda) =C_D\Lambda^{-0.5}$
in accordance with the analysis of the fixed point.

\begin{figure}[htbp]
\begin{center}
\includegraphics[width=0.6\hsize]{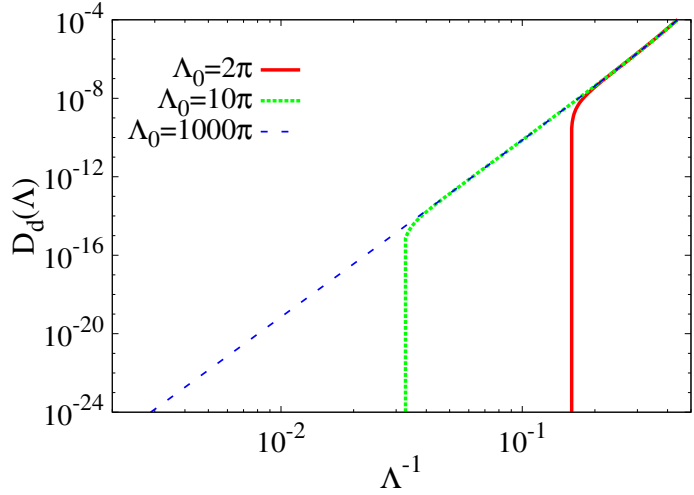}
\caption{
Graphs of $D_d(\Lambda)$ for $\pset^{KS}_0$ at $\Lambda_0 = 2\pi$, $10\pi$ and $1000\pi$.
\label{fig:Dd} }
\end{center}
\end{figure}

We note that $D_d(\Lambda)$ does not show the plateau region  in Fig.~\ref{fig:nuD-KS}.
However, this graph quickly converges to  $D_d(\Lambda)$
with $D_{d0}=0$ at $\Lambda_0=\infty$.
As shown in Fig.~\ref{fig:Dd}, the graphs of $D_d(\Lambda)$ with $D_{d0}=0$ at $\Lambda_0=2\pi$, $10\pi$ and $1000\pi$ do not exhibit the plateau.
Instead,  the graphs at $\Lambda_0=2\pi$ and $10\pi$ quickly approach  $D_d(\Lambda)$ in the limit $\Lambda_0 = \infty$ when $\Lambda$ is smaller than $\Lambda_0$.
Therefore, we define the bare parameter as $D_{dB}=0$  for such cases.

\subsection{definition of the most effective model}

Now, for the noisy KS equation with $\psetUV^{KS}$,
we consider the set ${\cal B}(\psetUV^{KS})$ of
bare parameters $\psetUV$, each of which
has the same  factors $C_\nu$, $C_D$, $C_K$, and $C_{D_d}$
in the universal range and
the same wavenumber scale  $\Lambda_{IR}$
as those for the noisy KS equation.
The graph of $\pset(\Lambda)$ for a given
$\psetUV \in {\cal B}(\psetUV^{KS})$ determines the wavenumber scale
$\Lambda_{UV}$ that represents the end of the ultraviolet plateau.
Note that
the value of $\Lambda_{UV}$ depends on $\psetUV \in {\cal B}(\psetUV^{KS})$.
Then, there is a special
model with $\psetUV \in {\cal B}(\psetUV^{KS})$ such that
$\Lambda_{UV}=\Lambda_{IR}^{KS}$. For this model,
as soon as the graph of $\pset(\Lambda)$ exits from the ultraviolet
plateau region, it enters the infrared universal range.
In other words, this special model represents the universal behavior of the
noisy KS equation in the most efficient manner. We refer to it
as {\it the most effective model for the universal range of
the noisy KS equation} with $\psetUV^{KS}$. Below,
we determine the most effective model.


\section{Representation of the parameter space}
\label{sec:parameterspace}
The solution trajectories for the RG equation are expressed as curves
in the five-dimensional parameter space consisting of
$\pset$.
We attempt to simplify a representation of the trajectories
so as to determine the most effective model.

First, recalling $\lambda(\Lambda) = \lambda_0$,
we may restrict the parameter space into the subspace
$\lambda=\lambda_0=1$.

\begin{figure}[tbp]
\begin{center}
\includegraphics[width=0.6\hsize]{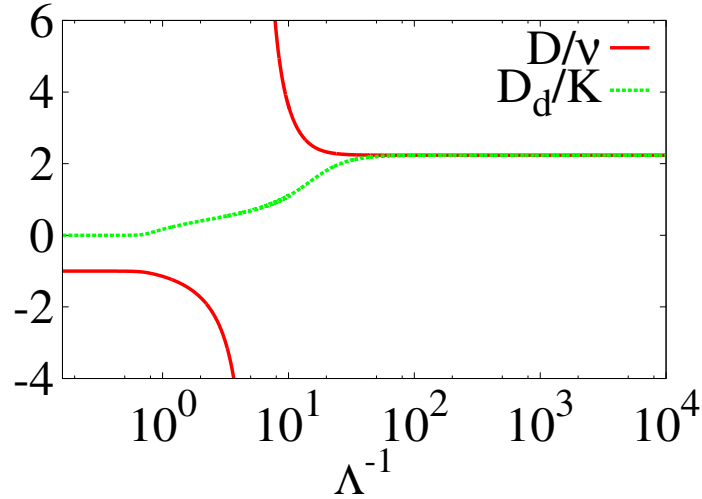}
\caption{ 
  Graphs
  of  $D(\Lambda)/\nu(\Lambda)$ and $D_d(\Lambda)/K(\Lambda)$ for $\pset_B^{KS}$.
  $D(\Lambda)/\nu(\Lambda)$ and $D_d(\Lambda) /K(\Lambda)$ converge
  to the same value, $ 2.24$.\label{fig:fdt-KS}}
\end{center}
\end{figure}

\begin{figure}[htbp]
\begin{center}
\includegraphics[width=0.8\hsize]{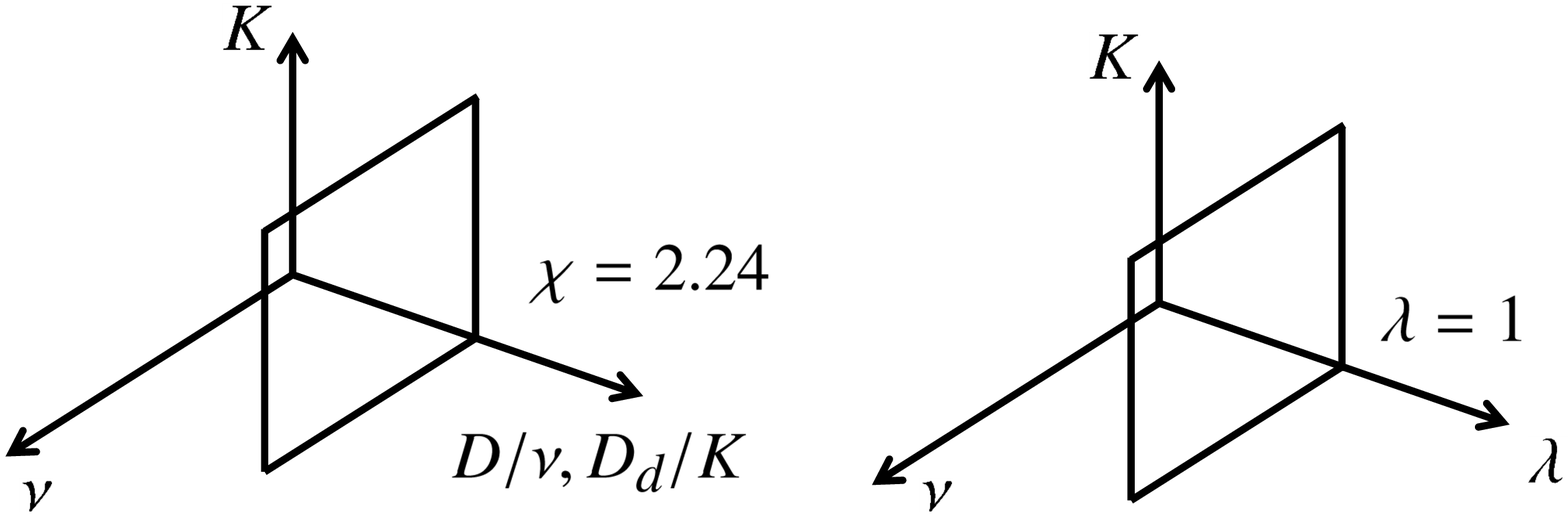}
\caption{
Two-dimensional subspace in the five-dimensional parameter space.
The solution trajectory of $\psetUV^{KS}$ is attracted to the subspace
defined by $\lambda=1$ and $D/\nu=D_d/K \equiv \chi=2.24$,
due to the tilt symmetry and the emergence of the time-reversal symmetry.
The most effective model should be defined on this plane. 
\label{fig:plane} }
\end{center}
\end{figure}

Next, as shown in Fig.~\ref{fig:fdt-KS}, we find
that $D(\Lambda)/\nu(\Lambda)$ and $D_d(\Lambda)/K(\Lambda)$
converge to the same value, 2.24,
in the universal range for the noisy KS equation. We can
explain this phenomenon as follows. First, for the generalized
KPZ equations with $\psetUV$ satisfying
$D_B/\nu_B=D_{dB}/K_B \equiv \chi >0 $,
we can show
the fluctuation-dissipation relation
with the effective temperature $\chi$ fixed by
using  a time-reversal symmetry.
The time-reversal transformation is given as
\begin{align}
h^{'}(k,\omega)&=-h(k,-\omega),\\
\tilde{h}^{'}(k,\omega)&=\tilde{h}(k,-\omega)-\frac{\nu_0 k^2}{D_0} h (k,-\omega).
\end{align}
The variation of the action (\ref{eq:action}) under this transformation is calculated as
\begin{align}
\delta S \equiv& S[h^{'}, \af^{'};\Lambda_0]-S[h,\af;\Lambda_0],\nonumber \\
=&\biggl(\frac{D_0}{\nu_0}-\frac{D_{d0}}{K_0}\biggr)\frac{\nu_0 K_0}{D_0}
\int \frac{d \omega d k}{(2\pi)^2 }\biggl(\frac{\nu_0}{D_0}k^2h(-k,-\omega) h(k, \omega)
-2\af(-k,-\omega) h(k, \omega)\biggr).
\end{align}
The generalized KPZ equation is invariant when $D_0/\nu_0=D_{d0}/K_0$ or $K_0=D_{d0}=0$.
This symmetry leads to the invariance property of $D(\Lambda)/\nu(\Lambda)$ and
$D_d(\Lambda)/K(\Lambda)$
along the solution trajectories of the RG equation.
See Appendix \ref{sec:WTid} for the time-reversal symmetry of the generalized KPZ equation and the derivation of the fluctuation-dissipation relation.
For the other cases where $D_B/\nu_B \not =D_{dB}/K_B$ including for
noisy KS equations, $D(\Lambda)/\nu(\Lambda)$ and $D_d(\Lambda)/K(\Lambda)$ change in $\Lambda$.
However, they satisfy $D(\Lambda)/\nu(\Lambda)= D_d(\Lambda)/K(\Lambda)$ in the universal range. Therefore, it is reasonable to conjecture that
the time-reversal symmetry emerges in the universal range.
Now, since the most effective model represents
the universal behavior most efficiently, this special model
should be in the subspace satisfying $D_B/\nu_B=D_{dB}/K_B \equiv \chi=2.24$.
On the basis of the results, we express the bare-parameter space
by $(\nu_B, K_B, D_B=2.24 \nu_B, D_{dB}=2.24K_B, \lambda_B=1)$, as illustrated in Fig.~\ref{fig:plane}.
For each value of $(\nu_B, K_B)$,  we have a model that exhibits the infrared universal
behavior of $\psetUV^{KS}$.


Finally, for  a generalized KPZ equation with $\psetUV$ at $\Lambda_0$
in the ultraviolet plateau region, we consider the following
scale transformation:
\begin{align}
X&=b_x x, \\ 
T&=b_t t,\\
H(X,T)&=b_h h(x,t),
\end{align}
which yields another generalized KPZ equation with a different collection
of bare parameters $\psetUV'$ at
$\Lambda_0'=b_x^{-1} \Lambda_0$ in the ultraviolet plateau region.
These are the equivalent models in different unit systems. 
For the cases that  $D=\chi \nu$ and $D_d=\chi K$, the equation for $H(X,T)$  is written as
\begin{align}
&\partial_T H = \nu^{'} \partial_X^2 H - K^{'} \partial_X^4 H +\frac{\lambda}{2}(\partial_X H)^2 + F,  \\
&\langle F(X,T) F(X^{'},T^{'}) \rangle = 2\chi^{'}(\nu^{'}-K^{'} \partial_X^2)\delta(T-T^{'})\delta(X-X^{'}),
\end{align}
where we have introduced
\begin{align}
\nu^{'}&=b_t^{-1}b_x^2\nu, \\
K^{'}&=b_t^{-1}b_x^4 K, \\
\lambda^{'}&=b_t^{-1}b_x^2 b_h^{-1}\lambda, \\
F(X,T)&=b_t^{-1}b_h \eta(x,t),  \\
\chi^{'}&=b_x^{-1}b_h^2 \chi.
\end{align}
By imposing $\chi'=\chi$ and
$\lambda'=\lambda$, we obtain 
 $b_h=b_x^{1/2}$ and $b_t=b_x^{3/2}$.
Then, we have the relation
\begin{align}
\nu_B^{'}&=b_x^{1/2}\nu_B, \\
K^{'}_B &=b_x^{5/2}K_B,
\end{align}
We find that $J \equiv K_B/\nu_B^5$ is invariant under the transformation. 
Thus, we parameterize $(\nu_B,K_B)$
as $(b_x^{1/2}, b_x^{5/2} J)$. The next problem is to determine
the values of $b_x$ and $J$ of the most effective model
for the universal range of the noisy KS equation.


\section{the most effective model}
\label{sec:MEmodel}
Since $J$ is invariant under the scale transformation, the determination
of $J$ can be separated from the determination of $b_x$.
Here, we notice the condition $\Lambda_{UV}=\Lambda_{IR}$ for
the most effective model. Because this condition is invariant under the scale transformation,
the value of $J$ is uniquely determined. Furthermore, the condition
$\Lambda_{IR}=\Lambda_{IR}^{KS}$ fixes the value of $b_x$. Below, we
explicitly calculate these values.


\begin{figure}[htbp]
  \begin{center}
    \begin{tabular}{c}
      \begin{minipage}{0.5\hsize}
        \begin{center}
          \includegraphics[width=\hsize]{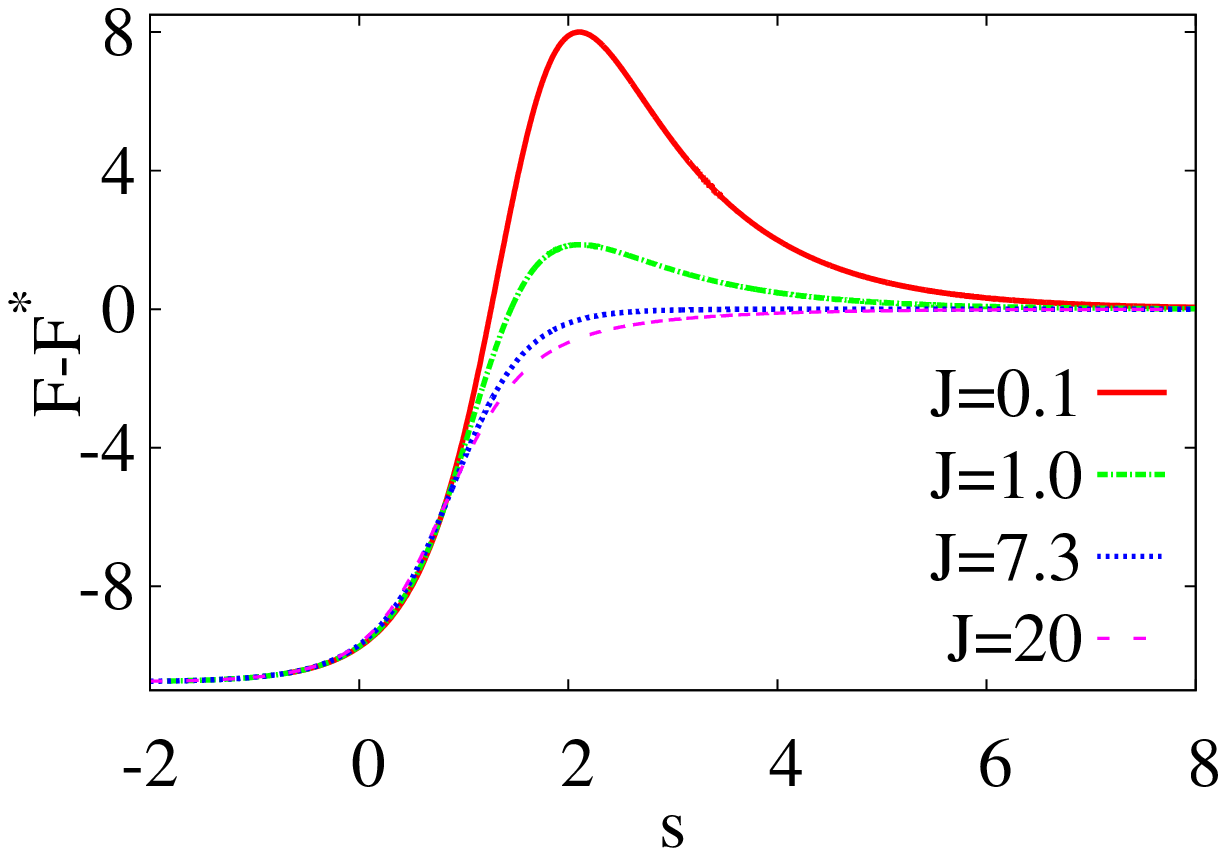}
        \end{center}
      \end{minipage}
      \begin{minipage}{0.5\hsize}
        \begin{center}
          \includegraphics[width=\hsize]{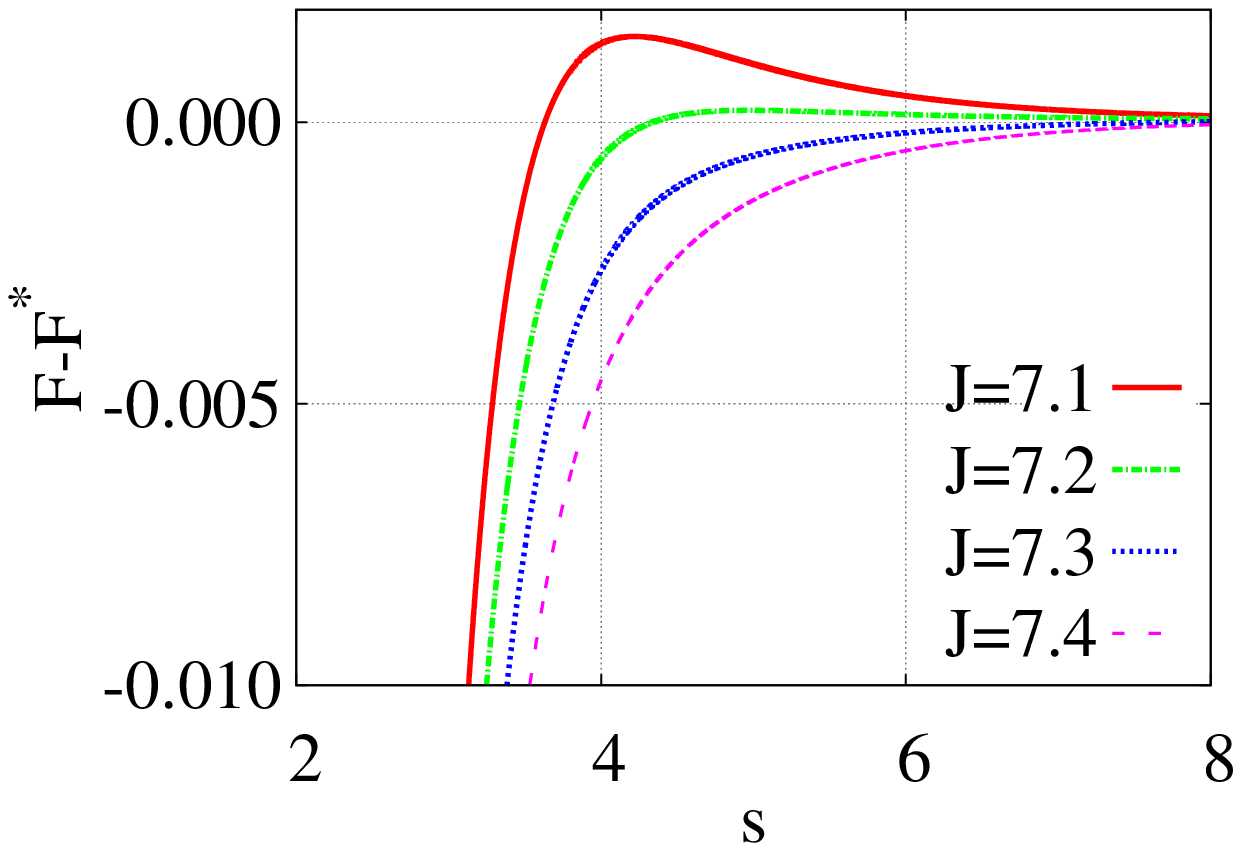}
        \end{center}
      \end{minipage}
    \end{tabular}
    \caption{
  Graphs of $F-F^*$ as a function of
 $s \equiv -\ln(J^{1/2} b_x\Lambda)$ for several $J$.}
    \label{fig:F-s}
  \end{center}
\end{figure}


In order to determine the value of $J$,
we study the dimensionless quantity $F(\Lambda) = \nu(\Lambda)/(K(\Lambda) \Lambda^2)$
as a function of
\begin{align}
s(\Lambda) \equiv -\ln(J^{1/2} b_x\Lambda), 
\end{align}
where $F$  and $s$ are invariant under the scale transformation. 
It should be noted that, for any $J$ and $b_x$, $F$ approaches
\begin{align}
F &\to e^{2s},
\end{align}
in the ultraviolet limit $s \to -\infty$, while
\begin{align}
F &\to F^* = 10.76,
\end{align}
in the infrared limit $s \to \infty$.
In Fig.~\ref{fig:F-s}, we show graphs of
$F$ as functions of $s$ for several values of $J$. In general, there
are two characteristic scales of $s$, the departure scale from
$e^{2s}$ and the relaxation scale to $F^*$, as clearly observed
for $J=0.1$. When $J$ increases, the peak of $F$ decreases
and eventually vanishes at $J=7.3$. In this case, the
transition scale between the infrared universal region and
the ultraviolet region is simply given by  the cross point $s_c$
of the ultraviolet behavior  $F=e^{2s}$ and the infrared behavior
$F^*=10.8$. 
That is, 
\begin{align}
e^{2s_c}=F^*,
\end{align}
which gives $s_c=1.2$.
Thus, we conclude that the value of $J$ of the most effective model
is $J=7.3$.

\begin{figure}[htbp]
  \begin{center}
    \begin{tabular}{c}

      \begin{minipage}{0.5\hsize}
        \begin{center}
          \includegraphics[width=\hsize]{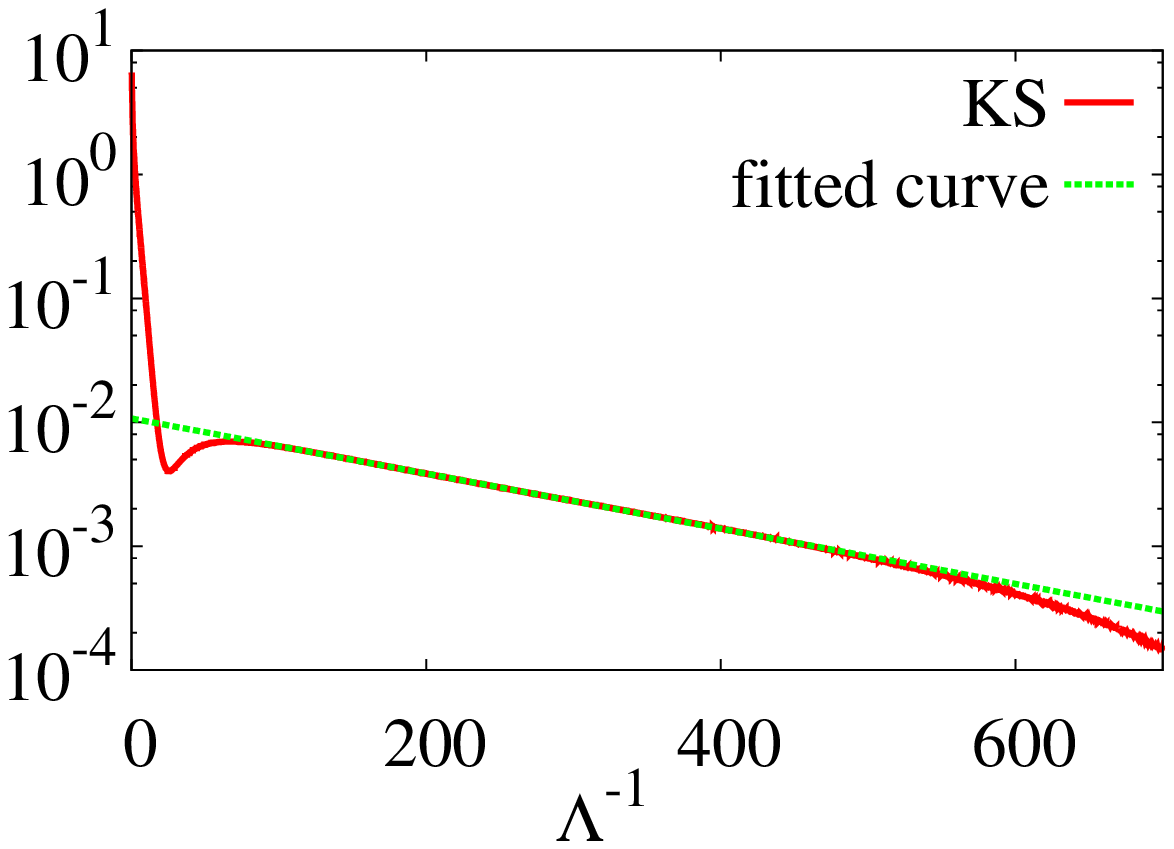}

        \end{center}
      \end{minipage}

      \begin{minipage}{0.5\hsize}
        \begin{center}
          \includegraphics[width=\hsize]{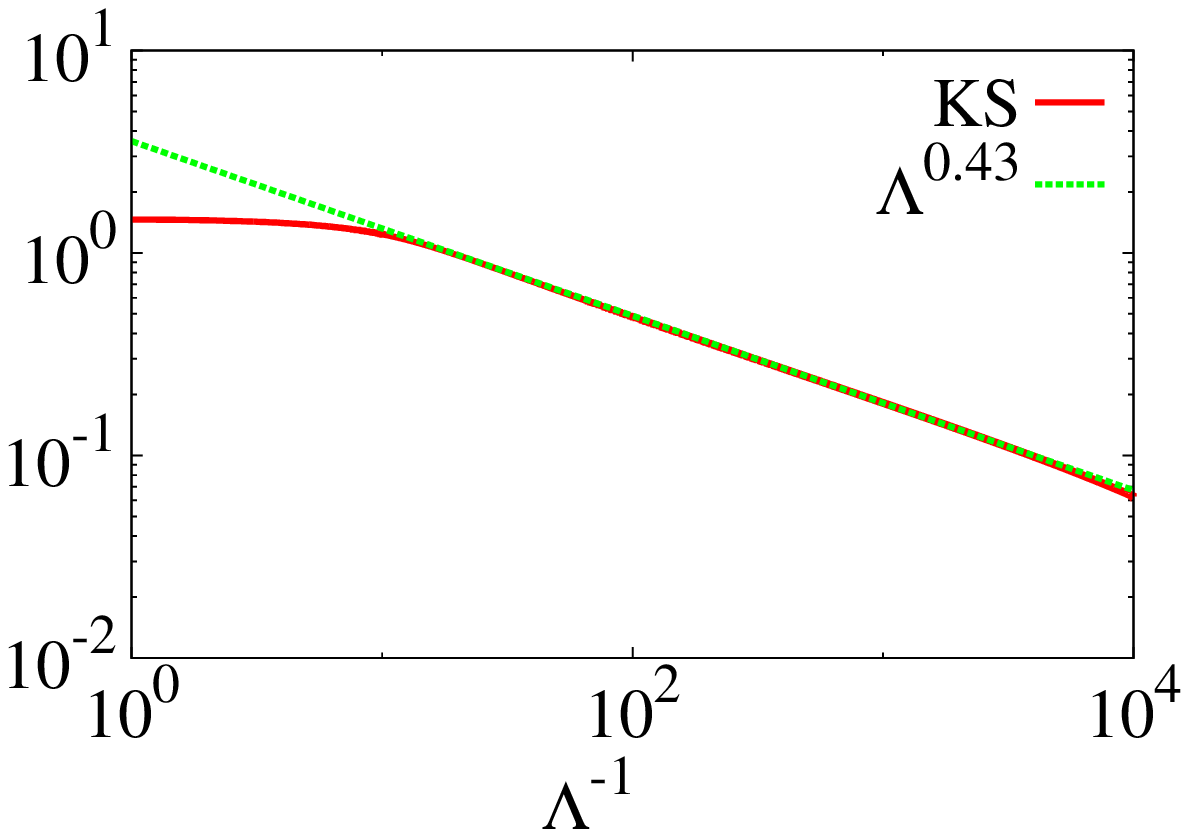}

        \end{center}
      \end{minipage}

    \end{tabular}
\caption{Graphs
  of $|\nu(\Lambda)-C_\nu\Lambda^{-1/2}+A\Lambda^B|$ for $\psetUV^{KS}$
  and the fitted curve in the left panel.
  The right panel shows the graphs of $|\nu(\Lambda)-C_\nu\Lambda^{-1/2}|$ 
  and $A \Lambda^B$ with $A=3.57$ and $B=0.431$.
  \label{fig:diff-nu}}
  \end{center}
\end{figure}

Next, we determine the value of $b_x$. From the cross point $s_c$,
we define the transition length scale
$\Lambda_c^{-1}$ by $s_c = -\ln(J^{1/2} b_x\Lambda_c)$,
which gives $\Lambda_c^{-1}= \sqrt{J F^*}b_x=8.9 b_x$.
Here, the value of $b_x$ is determined by identifying
$\Lambda_c$ with $\Lambda_{IR}^{KS}$. Then, we estimate
$\Lambda_{IR}^{KS}$ from the graph of $\nu(\Lambda)$
for the noisy KS equation under study. In Fig.~\ref{fig:diff-nu},
we show how $\nu(\Lambda)$ approaches $C_\nu\Lambda^{-0.5}$. We find
that $|\nu(\Lambda)-C_\nu\Lambda^{-0.5}| $ is well fitted to a power-law
function of $\Lambda^{-1}$, which does not provide any wavenumber scale.
Through more detailed analysis, we find a fitting function
\begin{align}
\nu(\Lambda)-C_\nu\Lambda^{-0.5} = -A\Lambda^{B}+C
\exp \biggl[ -\frac{\Lambda^{-1}}{D}\biggr],
\label{eq:difference}
\end{align}
with $A=3.57$, $B=0.431$, $C=1.1 \times 10^{-2}$, and $D=195$.
From the second term of (\ref{eq:difference}), we obtain
the characteristic scale $(\Lambda_{IR}^{KS})^{-1}=D=195$.
Now, from the condition
\begin{align}
(\Lambda_{IR}^{KS})^{-1}=\Lambda_c^{-1},
\end{align}
we obtain $b_x=22$. Thus, we have arrived at the most effective model
for the universal range of the noisy KS equation with $\psetUV^{KS}$,
where the collection of  the bare parameter values  of the most effective
model, $\pset_B^{ ME}$, is  determined as $(\nu_B=4.7,D_B=10, K_B=1.6 \times
10^4,D_{dB}=3.7\times 10^4, \lambda_B=1)$.

Now, the linear decay rate of the disturbance of a wavenumber $k$ in the
universal range is expressed as $\nu_B k^2+K_B k^4$ at an early time.
Here, we notice that $(\nu_B/K_B)^{0.5}$ defines
one wavenumber scale. Since the most effective model has only one
wavelength scale $\Lambda_c$, $(\nu_B/K_B)^{0.5} \simeq \Lambda_c$ holds.
This implies that the linear decay rate $\nu_B k^2+K_B k^4$ is
estimated as $\nu_B k^2$ for $k \ll \Lambda_c$. In this manner,
$\nu_B$ can be measured in experiments. Indeed, by applying this method to
the numerical simulation of the noisy KS equation, the result
$\nu_{B}^{\rm exp} \simeq 5.5$ was obtained \cite{PhysRevE.71.046138}.
Thus, our theoretical value $\nu_B  = 4.7 $ is in good agreement with the
numerical value.


\begin{figure}[htbp]
  \begin{center}
    \begin{tabular}{c}
      \begin{minipage}{0.5\hsize}
        \begin{center}
          \includegraphics[width=\hsize]{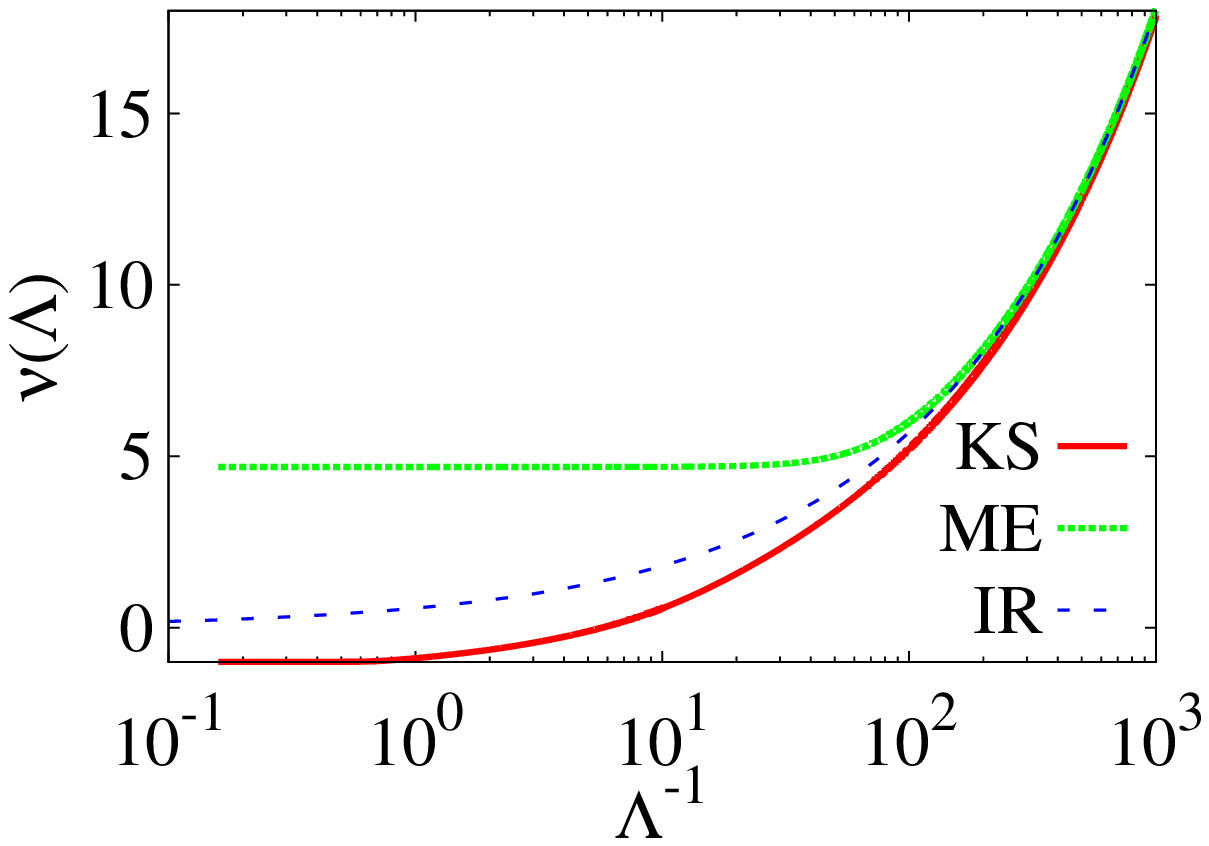}
        \end{center}
      \end{minipage}
      \begin{minipage}{0.5\hsize}
        \begin{center}
          \includegraphics[width=\hsize]{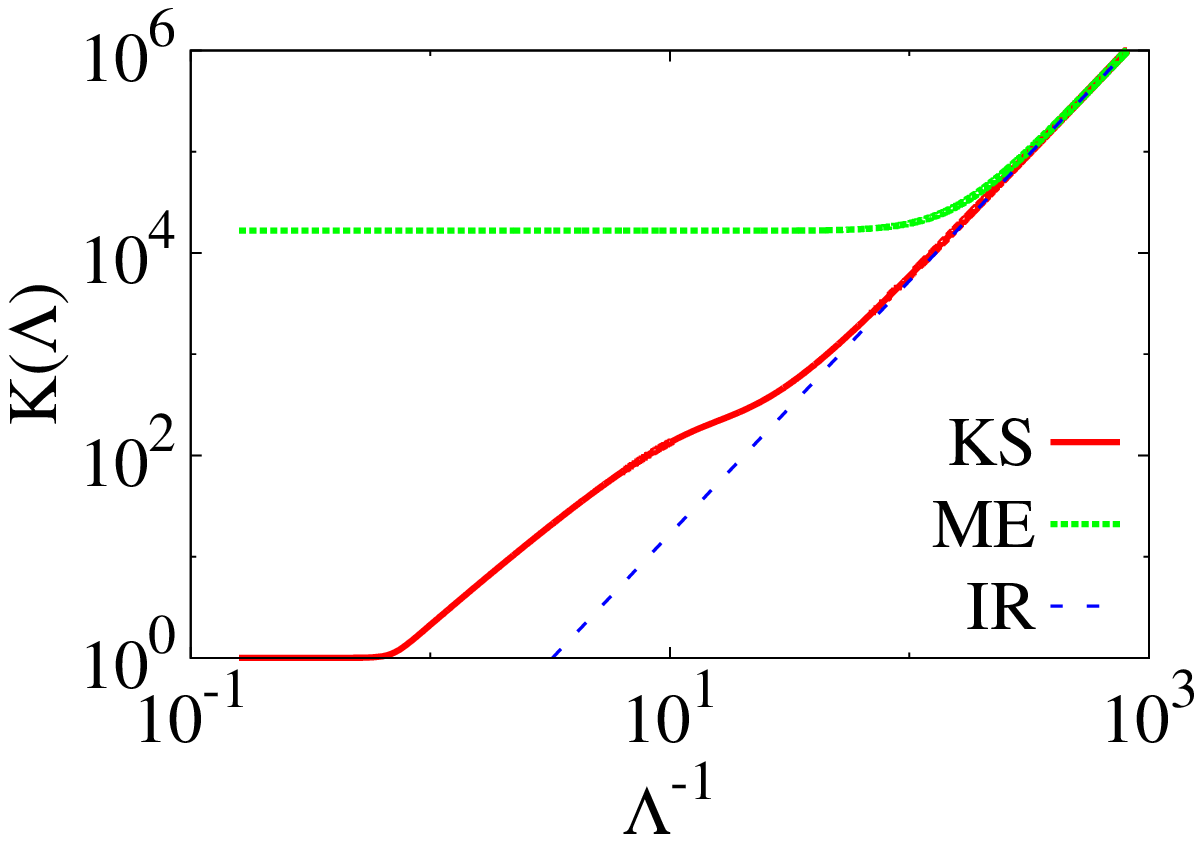}
        \end{center}
      \end{minipage}
    \end{tabular}
\caption{
Graphs of  $\nu(\Lambda)$ and $K(\Lambda)$
for $\psetUV^{KS}$, $\psetUV^{ME}$, and
the infrared scaling behaviors, respectively.
\label{fig:repredKPZ} }
  \end{center}
\end{figure}

\section{Concluding remarks} 
\label{sec:concluding}
The main result of this paper is illustrated in Fig. \ref{fig:repredKPZ}.
For a given noisy KS equation,
we construct the most effective model exhibiting
the same infrared universal behavior  with  just one cross-over
wavenumber scale $\Lambda_{IR}^{KS}$ connecting  the infrared behavior and the ultraviolet behavior.
We emphasize that our theory enables
us to calculate the bare surface tension $\nu_{B}$ of the effective
model in the universal range, which could not be obtained by previous studies.
  We conclude this paper by presenting
a few remarks.

\begin{figure}[htbp]
\begin{center}
\includegraphics[width=0.8\hsize]{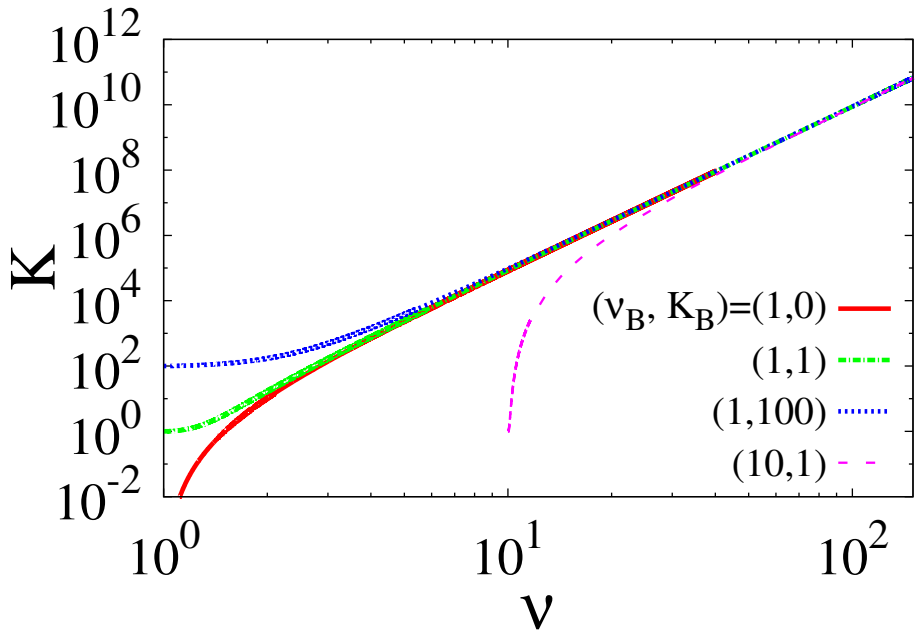}
\caption{Solution trajectories in the subspace of Fig.~\ref{fig:plane} for the bare parameters $(\nu_B,K_B)=(1,0)$, $(1,1)$, $(10,1)$ and $(1,100)$.
The trajectories converge to a single curve $K(\Lambda)=K^*(\nu(\Lambda), \chi, \lambda)$. 
We note that $K^*(\nu(\Lambda), \chi, \lambda)$ is a function of $\nu, \chi$ and $\lambda$, and does not explicitly depend on $\Lambda$.
Therefore, $\nu, \chi$ and $\lambda$ are relevant.
\label{fig:renormalizability} }
\end{center}
\end{figure}
The first remark is on the relevant parameter space in the universal range.
Since $\lambda(\Lambda)$ is a conserved quantity along
the solution of the RG equation,
it obviously depends on the initial condition $\pset_0$. Thus, it is relevant
in the universal range. Furthermore, $D/\nu- D_{d}/K$ is not relevant because $D(\Lambda)/\nu(\Lambda)-D_d(\Lambda)/K(\Lambda)$ approaches zero.
At the same time, $\chi=D/\nu$ is a relevant parameter because
its value is invariant along the solution trajectory  when $D_0/\nu_0=D_{d0}/K_0$.
Finally, in the limit $\Lambda \to 0$, $K(\Lambda) \Lambda^2/ \nu(\Lambda)$
approaches the universal constant value $0.0929$ which is independent
of $\pset_0$. Thus, we can state that $K(\Lambda) \Lambda^2/ \nu(\Lambda)$
is irrelevant, following the argument in \cite{polchinski1984,Weingberg1995}.
In other words, $\nu(\Lambda)$ and $K(\Lambda)$ are not independent of
each other in the universal range, as shown in Fig.~\ref{fig:renormalizability}. 
In summary, the relevant parameter
space in the universal range is spanned by the three parameters
$(\nu, \chi, \lambda)$. However, the parameter $K$ cannot
be negligible because the irrelevant parameter
$K(\Lambda) \Lambda^2/ \nu(\Lambda)$ is not zero in the universal range.
This is different from many standard  RG analysis \cite{Wilson1974}.

Second, we remark that the original Yakhot conjecture claims
a statistical property of the deterministic KS equation \cite{PhysRevA.24.642}.
Here, we discuss the noiseless limit $D_0 \to 0$ for the noisy
KS equation. In this case, we obtain $\chi \to 0$ which is not consistent
with observations. This implies that
the lowest order contribution in loop expansions is not sufficient
to yield statistical properties for the small $D_0$ limit.
In order to overcome this situation, we have to formulate a non-perturbative
calculation. This is an interesting problem for future work.

Finally, we expect that the concept proposed in this paper
will be applied to various systems, although we have studied a specific
phenomenon as an example of scale-dependent parameters.  The most
interesting example may be fluid turbulence. The effective model
for the universal range in the turbulence is given by a noisy Navier-Stokes
equation \cite{Dominicis,Fournier-Frisch1978,Fournier-Frisch1983,Yakhot-Orszag,Yakhot-Orszag2,Yakhot-Smith,Eyink}
\begin{align}
\partial_t \bm{v}(\bm{x},t) &= -\nabla p(\bm{x},t) + \eta_0 \nabla^2 \bm{v}(\bm{x},t)+\bm{f}(\bm{x},t), \\
\langle f_i (\bm{x},t) f_j(\bm{x}^{'}, t^{'}) \rangle &= 2P_{i j}(\nabla) F(\bm{x}-\bm{x}^{'})\delta(t-t^{'}),
\end{align}
where $\bm{v}$ is the fluid velocity, $p$ is the pressure, $\eta_0$ is the viscosity and $\bm{f}$ is the noise. 
Here, $P_{i j}$ and $F$ in the Fourier space are given as 
\begin{align}
P_{i j}(\bm{k}) &=\delta_{i j}-\frac{k_i k_j}{k^2}, \\
F(\bm{k}) &=(2\pi)^d 2 D_0  k^{-y},
\end{align}
where $d$ is the space dimension, $D_0$ is the noise strength, and $y$ is a positive parameter.
When $y=d=3$, this model exhibits the Kolmogorov scaling law
\begin{align}
E(k)= C k^{-5/3},
\end{align}
where $E(k)$ is the energy spectra and  $C$ is a universal constant and takes the value $\sim 1.5$.
The analysis of solution trajectories of an RG equation for
such the noisy Navier-Stokes equation may provide
 fresh insight into the understanding of the turbulence such as the universal constant $C$. We hope that
this paper stimulates the study of whole solutions of  RG equations
in various research fields.

The authors thank K.~A.~Takeuchi, M.~Itami, and T.~Haga for useful discussions.
The present study was supported by KAKENHI (Nos. 25103002 and 17H01148).



\appendix


\section{Ward-Takahashi identities}
\label{sec:WTid}
In this section, we prove
\begin{equation}
  \lambda(\Lambda)=\lambda_0, \label{eq:lambda}
\end{equation}
for all generalized KPZ equations, and
\begin{equation}
\frac{\nu(\Lambda)}{D(\Lambda)} =  \frac{\nu_0}{D_0}, \label{eq:nu/D}
\end{equation}
for $K_0=D_{d0}=0$ or $K_0/D_{d0}=\nu_0/D_0$, and
\begin{equation}
\frac{K(\Lambda)}{D_d(\Lambda)} =  \frac{\nu_0}{D_0}, \label{eq:K/Dd}
\end{equation}
for $K_0/D_{d0}=\nu_0/D_0$.
These results are easily obtained from the following Ward-Takahashi identities \cite{freyPRE1994,Canet2011}:
\begin{equation}
(G^{-1})_{\tilde{h} h}(k=0, \omega; \Lambda) = -i \omega, \label{eq:WTid-shift1}
\end{equation}
\begin{equation}
  i \lambda_0 k_1 \partial_{\omega_1} (G^{-1})_{\tilde{h} h}(k_1,\omega_1;\Lambda)=
\lim_{\omega, k \to 0}\partial_k \Gamma_{\tilde{h} h h}(k_1,\omega_1;k,\omega; \Lambda),
\label{eq:WTid-tilt}
\end{equation}
and
\begin{equation}
 G_{\tilde{h}h}^{-1}(k_1, \omega_1;\Lambda)+G_{\tilde{h}h}^{-1}(-k_1, -\omega_1;\Lambda)
=-\frac{\nu_0 k_1^2}{D_0} G_{\tilde{h}\tilde{h}}^{-1}(k_1, -\omega_1;\Lambda).
 \label{eq:fdt1}
\end{equation}
These identities are relaed to invariance properties of the MSRJD action
for a shift transformation, a tilt transformation, and a time-reversal transformation, respectively.
In the next subsections, we will derive (\ref{eq:WTid-shift1})-(\ref{eq:fdt1}) following the arguments \cite{freyPRE1994,Canet2011}.

Here, we derive (\ref{eq:lambda})-(\ref{eq:K/Dd}) from (\ref{eq:WTid-shift1})-(\ref{eq:fdt1}).
First, by differentiating (\ref{eq:WTid-tilt}) with respect to $k_1$ and taking the limit $k_1 \rightarrow 0$, we have
\begin{align}
i\lambda_0 \partial_{\omega_1}(G^{-1})_{\tilde{h} h}(k_1=0,\omega_1;\Lambda)= \lim_{\omega, k,k_1 \to 0}\partial_k \partial_{k_1}\Gamma_{\tilde{h} h h}(k_1,\omega_1;k,\omega; \Lambda).
\label{eq:WTid-tilt2}
\end{align}
Next, we substitute (\ref{eq:WTid-shift1}) to (\ref{eq:WTid-tilt2}) and take the limit $\omega_1 \rightarrow 0 $.
Then, we obtain
\begin{align}
\lambda_0=\lim_{\omega, \omega_1,k,k_1 \to 0}\partial_k \partial_{k_1}\Gamma_{\tilde{h} h h}(k_1,\omega_1;k,\omega; \Lambda).
\end{align}
By recalling the definition (\ref{eq:definition of lambda}),  we find that this equality is (\ref{eq:lambda}).
Second, we differentiate (\ref{eq:fdt1})  twice with respect to $k_1$.
Then, we have
\begin{align}
\partial_{k_1}^2 G_{\tilde{h}h}^{-1}(k_1, \omega_1;\Lambda)+\partial_{k_1}^2G_{\tilde{h}h}^{-1}(-k_1, -\omega_1;\Lambda)
=
-\frac{\nu_0}{D_0} ( 2 + 2k_1\partial_{k_1}+ k_1^2\partial_{k_1}^2 )G_{\tilde{h}\tilde{h}}^{-1}(k_1, -\omega_1;\Lambda).
\end{align}
By taking the limit $\omega_1, k_1 \rightarrow 0 $ and using (\ref{eq:definition of nu}) and (\ref{eq:definition of D}), we obtain (\ref{eq:nu/D}).
Finally, by differentiating (\ref{eq:fdt1}) four times with respect to $k_1$,  we arrive at (\ref{eq:K/Dd}).

\subsection{Proof of (\ref{eq:WTid-shift1})}

We consider a  shift transformation
\begin{align}
h^{'}(x,t)=h(x,t)+c(t),
\end{align}
where $c(t)$ is an infinitesimal parameter that depends on time.
The variation of the  MSRJD action for the transformation is calculated
as
\begin{align}
  S[h^{'},\af^{'};\Lambda_0] - S[h,\af;\Lambda_0]=
  \int dt dx \af(x,t) \partial_t c(t).
  	\label{eq:variation of shift}
\end{align}
It should be noted that this simple form comes from the invariance property
of the MSRJD action for the time-independent $c$~\footnote{In general, by assuming time dependence of the infinitesimal parameter for a continuous symmetry transformation,
we can obtain non-trivial identities such as (\ref{eq:WTid-shift1}).
This technique, which has been referred to as ``gauging a global symmetry'', is standard when we  derive  identities from a continuous  global symmetry~\cite{Weingberg1995}.
For such a case,    the variation of an action under a time-gauged transformation is expressed as
$\delta S = \int dt Q(t) \partial_t \epsilon(t)$,
where $Q(t)$ is a Noether charge of the corresponding global symmetry, and $\epsilon(t)$ is the time-gauged infinitesimal parameter.
The Noether charge of the shift symmetry is calculated as $Q_{\rm shift} = \int dx \af(x,t)$, which is consistent with (\ref{eq:variation of shift}).}.
Then, the variation of the effective MSRJD action is derived as
\begin{align}
S[h^{<'},\af^{'<};\Lambda]=&-\log \int \mathcal{D} [h^{>'}, \af^{>'}]\exp\biggl[-S[h^{'},\af^{'}; \Lambda_0]\biggr],
\nonumber \\
=&-\log \int \mathcal{D} [h^{>}, \af^{>}]\exp\biggl[-S[h,\af; \Lambda_0]-\int dtdx \af(x,t) \partial_t c(t) \biggr], \nonumber \\
=&\int dtdx \af^<(x,t) \partial_t c(t) -\log \int \mathcal{D} [h^{>}, \af^{>}]\exp\biggl[-S[h,\af; \Lambda_0]-\int dtdx \af^>(x,t) \partial_t c(t) \biggr],
\nonumber \\
=&S[\eh,\eaf; \Lambda]+\int dtdx \af^<(x,t) \partial_t c(t).
\label{eq:variation}
\end{align}
When we obtain the fourth line in (\ref{eq:variation}) from the third line,
we have used
\begin{align}
\int dtdx \af^>(x,t) \partial_t c(t)
 &= \int dt \partial_t c(t) \biggl( \int dx \af^>(x,t) \biggr), \nonumber \\
 &= \int dt \partial_t c(t) \af^>(k=0,t), \nonumber \\
 &=0.
\end{align}
Here, noting the trivial relation
\begin{equation}
S[h^{<'},\af^{'<};\Lambda]=S[\eh,\eaf; \Lambda]+
\int dt dx \frac{\delta S[\eh,\eaf; \Lambda]}{\delta \eh(x,t)}c(t),
\end{equation}
we rewrite (\ref{eq:variation}) as
\begin{align}
\int dt dx \biggl( \frac{\delta S[\eh,\eaf; \Lambda]}{\delta \eh(x,t)}c(t)-\eaf(x,t) \partial_t c(t) \biggr)=0,
\end{align}
which is further expressed as
\begin{align}
\int dt dx \biggl( \frac{\delta S[\eh,\eaf; \Lambda]}{\delta \eh(x,t)}+\partial_t\eaf(x,t)  \biggr)c(t)=0.
\end{align}
Since this equality holds for any $c(t)$,  we obtain
\begin{align}
\int dx \biggl( \frac{\delta S[\eh,\eaf; \Lambda]}{\delta \eh(x,t)}+\partial_t\eaf(x,t)  \biggr) =0.
\label{eq:WTid-shift2}
\end{align}
The differentiation of (\ref{eq:WTid-shift2}) with respect
to $\eaf(t^{'},x^{'})$ leads to
\begin{align}
  \int dx \biggl( (G^{-1})_{\tilde{h} h}(x^{'}-x, t^{'}-t; \Lambda)
  +\partial_t \delta (t-t^{'})\delta(x-x^{'})\biggr)=0.
\end{align}
By performing the Fourier transformation,
we arrive at (\ref{eq:WTid-shift1}).

\subsection{Proof of (\ref{eq:WTid-tilt}) }

We consider a tilt transformation
\begin{align}
h^{'}(x,t)&=h(x+\lambda_0 v t,t)+v x,  \\
\tilde{h}^{'}(x,t)&=\tilde{h}(x+\lambda_0 v t,t),
\end{align}
where $v$ is an infinitesimal parameter.
The tilt transformation for their Fourier transforms is expressed as
\begin{align}
h^{'}(k,t)&=e^{i\lambda_0 v k t}h(k,t)-i v\partial_k \delta (k),  \\
\af^{'}(k,t)&=e^{i\lambda_0 v k t}h(k,t).
\end{align}
We then find the symmetry property
\begin{align}
S[h^{<'}+h^{>'},\af^{<'}+\af^{>'}; \Lambda_0]=S[h^<+h^>,\af^<+\af^>; \Lambda_0],
\end{align}
from which we obtain
\begin{align}
S[h^<,\af^<; \Lambda]&=-\log \int \mathcal{D} [h^>, \af^>]\exp \biggl[-S[h^{<}+h^{>},\af^{<}+\af^{>}; \Lambda_0]\biggr], \nonumber \\
&=-\log \int \mathcal{JD} [h^{>'}, \af^{>'}]\exp[-S\biggl[h^{<'}+h^{>'},\af^{<'}+\af^{>'}; \Lambda_0]\biggr], \nonumber \\
&=S[h^{<'},\af^{<'}; \Lambda]-\log\mathcal{J}, \label{eq:TRaction}
\end{align}
where $\mathcal{J} =1+ v a$ is the Jacobian for the tilt transformation, and
$a$ is a field independent quantity.
The expansion of (\ref{eq:TRaction}) in $v$ leads to the identity
\begin{align}
\int dk dt \biggl[ i\lambda_0 kt \biggl( \frac{\delta S[\eh,\eaf; \Lambda]}{\delta \eh(k,t)} \eh(k,t)
+\frac{\delta S[\eh,\eaf; \Lambda]}{\delta \eaf(k,t)} \eaf(k,t)\biggr)
+i \delta(k)\partial_k \frac{\delta S[\eh,\eaf; \Lambda]}{\delta \eh(k,t)}
-a \biggr]=0.
\end{align}
We differentiate this identity with respect to $\eaf(k_1,t_1)$ and $\eh(k_2,t_2)$.
Then, we have
\begin{align}
\int dk dt \biggl[ i\lambda_0 kt&
\biggl(
\frac{\delta^2 S[\eh,\eaf; \Lambda]}{\delta \eh(k,t)\delta \eaf(k_1,t_1)} \delta(k_2-k)\delta(t_2-t)
+\frac{\delta^3 S[\eh,\eaf; \Lambda]}{\delta \eh(k,t)\delta \eaf(k_1,t_1) \delta \eh(k_2,t_2) } \eh(k,t)
  \nonumber \\
+&\frac{\delta^2 S[\eh,\eaf; \Lambda]}{\delta \eaf(k,t) \delta \eh(k_2,t_2)} \delta(k_1-k)\delta(t_1-t)
+\frac{\delta^3 S[\eh,\eaf; \Lambda]}{\delta \eh(k,t)\delta \eaf(k_1,t_1)\delta \eh(k_2,t_2)} \eaf(k,t)
\biggr) \nonumber \\
&+i \delta(k)\partial_k \frac{\delta^3 S[\eh,\eaf; \Lambda]}{\delta \eh(k,t)\delta \eaf(k_1,t_1) \delta \eh(k_2,t_2)}
 \biggr]=0.
\end{align}
By taking the limit $\eaf,\eh \rightarrow 0$ and recalling the definitions given in (\ref{eq:def of propagator}) - (\ref{eq:def of vertex}), we obtain
\begin{align}
\lambda_0(k_1t_1+k_2t_2) &(G^{-1})_{\tilde{h} h}(k_1,t_1-t_2;\Lambda) \delta (k_1+k_2)
\nonumber \\
&=-i \lim_{k \to 0}\partial_k \int dt  \Gamma_{\tilde{h} h h}(k,k_1,k_2;t-t_1, t_2-t_1; \Lambda ) \delta (k+k_1+k_2).
\end{align}
The Fourier transform of this equality is (\ref{eq:WTid-tilt}).

\subsection{Proof of (\ref{eq:fdt1})}
We consider a time-reversal transformation
\begin{align}
h^{'}(k,\omega)&=-h(k,-\omega),\\
\tilde{h}^{'}(k,\omega)&=\tilde{h}(k,-\omega)-\frac{\nu_0 k^2}{D_0} h (k,-\omega).
\end{align}
The variation of the action (\ref{eq:action}) under this transformation is calculated as
\begin{align}
\delta S \equiv& S[h^{'}, \af^{'};\Lambda_0]-S[h,\af;\Lambda_0],\nonumber \\
=&\biggl(\frac{D_0}{\nu_0}-\frac{D_{d0}}{K_0}\biggr)\frac{\nu_0 K_0}{D_0}
\int \frac{d \omega d k}{(2\pi)^2 }\biggl(\frac{\nu_0}{D_0}k^2h(-k,-\omega) h(k, \omega)
-2\af(-k,-\omega) h(k, \omega)\biggr).
\end{align}
The generalized KPZ equation is invariant when $D_0/\nu_0=D_{d0}/K_0$ or $K_0=D_{d0}=0$.

Here, we focus on the case $D_0/\nu_0=D_{d0}/K_0$.
Then, we obtain
\begin{align}
 S[h^{<'},\af^{<'};\Lambda ]=S[h^<,\af^<;\Lambda]-\log\mathcal{J},
\end{align}
where $\mathcal{J}$ is the Jacobian of the time-reversal transformation.
By differentiating this equality with respect to $\eaf(k_1, \omega_1)$ and $\eh(k_2,\omega_2)$, we have
\begin{align}
\frac{\delta^2 S[\eh,\eaf; \Lambda]}{\delta(\eaf(k_1, \omega_1))\delta(\eh(k_2, \omega_2))}
=&-\frac{\nu_0 k_1^2}{D_0}\frac{\delta^2 S[h^{<'},\af^{<'}; \Lambda]}{\delta (\af^{<'}(k_1, -\omega_1))\delta(\af^{<'}(k_2, -\omega_2))}
\nonumber \\
 &-\frac{\delta^2 S[h^{<'}, \af^{<'}; \Lambda]}{\delta (h^{<'}(k_1, -\omega_1))\delta(\af^{<'}(k_2, -\omega_2))}, 
\end{align}
where we have used the relation
\begin{align}
\frac{\delta}{\delta h(k,\omega)}&=
 -\frac{\delta}{\delta h^{'}(k,-\omega)}-\frac{\nu_0 k^2}{D_0} \frac{\delta}{\delta  \af^{'}(k,-\omega)}, \\
 \frac{\delta}{\delta \af(k,\omega)} &= \frac{\delta}{\delta \af^{'}(k,-\omega)}.
\end{align}
By recalling the definition given in (\ref{eq:def of propagator}) - (\ref{eq:def of vertex}), we obtain
\begin{align}
 G_{\tilde{h}h}^{-1}(k_1, \omega_1;\Lambda)&\delta(\omega_1+\omega_2)\delta(k_1+k_2)
 \nonumber \\
=&-\biggl(\frac{\nu_0 k_1^2}{D_0} G_{\tilde{h}\tilde{h}}^{-1}(k_1, -\omega_1;\Lambda)
 +G_{\tilde{h}h}^{-1}(-k_1, -\omega_1;\Lambda)\biggr)\delta(\omega_1+\omega_2)\delta(k_1+k_2).
 \label{eq:fdt2}
\end{align}
By rearranging (\ref{eq:fdt2}), we arrive at the identities (\ref{eq:fdt1}).

\end{document}